\documentclass[twocolumn]{aastex631}
\usepackage{CJK}
\usepackage{upgreek}
\usepackage{amsmath}
\usepackage{booktabs}

\date{\today}

\submitjournal{\apj}

\shorttitle{GRB in Binary System}
\shortauthors{Z.-C. Zou, B.-B. Zhang, Y.-F. Huang, \& X.-H. Zhao}

\begin{document}
\begin{CJK*}{UTF8}{gbsn}

\title{Gamma-Ray Burst in a Binary System}

\correspondingauthor{Ze-Cheng Zou}
\email{zou.ze-cheng@smail.nju.edu.cn}

\correspondingauthor{Bin-Bin Zhang}
\email{bbzhang@nju.edu.cn}

\correspondingauthor{Yong-Feng Huang}
\email{hyf@nju.edu.cn}

\author[0000-0002-6189-8307]{Ze-Cheng Zou (邹泽城)}
\affiliation{School of Astronomy and Space Science, Nanjing University, Nanjing 210023, China}

\author[0000-0003-4111-5958]{Bin-Bin Zhang}
\affiliation{School of Astronomy and Space Science, Nanjing University, Nanjing 210023, China}
\affiliation{Key Laboratory of Modern Astronomy and Astrophysics (Nanjing University), Ministry of Education, Nanjing 210023, China}
\affiliation{Department of Physics and Astronomy, University of Nevada Las Vegas, NV 89154, USA}
\author[0000-0001-7199-2906]{Yong-Feng Huang (黄永锋)}
\affiliation{School of Astronomy and Space Science, Nanjing University, Nanjing 210023, China}
\affiliation{Key Laboratory of Modern Astronomy and Astrophysics (Nanjing University), Ministry of Education, Nanjing 210023, China}

\author[0000-0003-3659-4800]{Xiao-Hong Zhao (赵晓红)}
\affiliation{Yunnan Observatories, Chinese Academy of Sciences, 650216, Kunming, China}
\affiliation{Center for Astronomical Mega-Science, Chinese Academy of Sciences, Beijing, China}

\begin{abstract}
Regardless of their different types of progenitors and central engines,
gamma-ray bursts (GRBs) were always assumed to be standalone systems after they
formed. Little attention has been paid to the possibility that a stellar
companion can still accompany a GRB itself. This paper investigates such a
GRB-involved binary system and studies the effects of the stellar companion on
the observed GRB emission when it is located inside the jet opening angle.
Assuming a typical emission radius of $\sim10^{15}\,$cm, we show
that the blockage by a companion star with a radius of
$R_\mathrm{c}\sim67\,\mathrm{R_\odot}$ becomes non-negligible when it is located
within a typical GRB jet opening angle (e.g., $\sim10$ degrees) and beyond the
GRB emission site. In such a case, an on-axis observer will see a GRB with a
similar temporal behavior but 25\% dimmer. On the other hand, an off-axis
observer outside the jet opening angle (hence missed the original GRB) can see a
delayed ``reflected" GRB, which is much fainter in brightness, much wider in the
temporal profile and slightly softer in energy. Our study can naturally explain
the origin of some low-luminosity GRBs. Moreover, we also point out that the
companion star may be shocked if it is located inside the GRB emission site,
which can give rise to an X-ray transient or a GRB followed by a delayed X-ray
bump on top of X-ray afterglows.
\end{abstract}


\section{Introduction} \label{sec:intro}
\end{CJK*}

Gamma-ray bursts (GRBs) represent the most high-energy transients in the
universe, with the isotropic bolometric emission energy, $E_\mathrm{iso}$,
ranging from $\sim 10^{46}\,\mathrm{erg}$ to $\sim 10^{55}\,\mathrm{erg}$
\citep{2018pgrb.book.....Z}. The prompt emission of a GRB is believed to be
generated by a relativistic jet launched from the central engine, while the
afterglow is likely produced by external shocks from the interaction between
the jet material and circumburst medium \citep[for a review, see][]{2015PhR...561....1K}.

GRBs are traditionally classified into long or short events according to
their durations \citep{1993ApJ...413L.101K}.
The progenitors of short GRBs have long been believed to be
mergers involving neutron stars
\citep[e.g.,][]{1989Natur.340..126E,1992ApJ...395L..83N,2007PhR...442..166N}
and are finally confirmed through GW170817/GRB
170817A association \citep{2017PhRvL.119p1101A,2017ApJ...848L..13A}.
It is interesting to note that neutron stars may actually be
strange quark stars \citep{2021arXiv210304165G}.
On the other hand, long GRBs are identified as dying massive stars
\citep{1997Natur.386..686V,1997Natur.387..783C,1997Natur.389..261F},
among which the most popular candidates are Wolf-Rayet (WR) stars
\citep{1993ApJ...405..273W}. In the so-called ``collapsar'' model, the core of a
dying WR star is required to rotate very fast to launch a relativistic jet
\citep{1999ApJ...524..262M}, though how to spin it up has been long debated
\citep[for a review, see][]{2016SSRv..202...33L}.

Many stars are in binary or even multiple systems
\citep{2013ARA&A..51..269D,2018ApJ...854..147B}. Therefore, one may
naturally expect that the progenitors of GRBs can be part of such systems.
Binary star progenitors of long GRBs have been previously investigated by
many authors
\citep[e.g.,][]{2004MNRAS.348.1215I,2005ApJ...623..302F,2005A&A...435..247P,2006ApJ...637..914W,2007A&A...465L..29C,2009A&A...497..243D}.
The binary interaction can significantly change the characteristics of the GRB
progenitor. In this so-called binary star scenario, a massive star is spun up by
its close companion via mass transfer and/or tidal forces to produce a GRB
\citep{2005ApJ...623..302F,2006ApJ...637..914W,2007A&A...465L..29C}. It is
interesting to note that a massive WR star in a binary system was recently
reported \citep{2019NatAs...3...82C}, which may potentially provide a
long-duration GRB progenitor. On the other hand, the massive star can also be in
a wide binary, where the system has a weak interaction. Hence, the progenitor is
in fact an effectively single star, which can generate a GRB via chemically
homogeneous evolution
\citep[e.g.,][]{2005A&A...443..643Y,2006ApJ...637..914W,2012A&A...542A.113Y}.

On the other hand, a GRB is always assumed to be a standalone system
after it is formed. The prompt emission of GRBs is usually thought to be of
internal origin, so the central engine itself determines most of the observed
properties. In this scenario, the existence of a stellar companion, if any, is
ignored. Some early efforts
\citep[e.g.,][]{1983ApJ...275L..59L,1985Natur.314..242R,1991ApJ...370..341D} have
been made to predict that the reprocessing of the original GRB photons by nearby
stars can produce an optical or X-ray counterpart with a time scale ranging from
$\sim1\,$s to $\sim10^3\,$s. However, the case that a GRB is produced with a
stellar companion located inside the jet opening angle has never been studied.
In this paper, we will discuss such scenarios in detail. As shown
below, the companion can occult or reflect the emission of the GRB,
depending on the viewing angle of the observer. The GRB light curve and
spectrum will therefore be modified.

This paper studies in detail the possible effects when a GRB happens in a
binary system. Two types of effects, namely occultation and reflection, are
proposed and investigated. In Section~\ref{sec:binary}, we describe the basic
framework of the binary system. The observational effects of the stellar
companion are presented in Section~\ref{sec:obs}. Finally, we briefly
conclude and discuss our results in Section~\ref{sec:conclude}. Throughout
this paper, a flat $\Lambda$CDM cosmology with
$H_0=67.4\,\mathrm{km\,s^{-1}\,Mpc^{-1}}$ and $\Omega_\mathrm{m}=0.315$ is
adopted \citep{2020A&A...641A...6P}.

\section{Binary Configuration} \label{sec:binary}

In this paper, the term ``binary'' refers to a system consisting of a
primary object producing a GRB and its stellar companion, i.e., we do not
specify a certain progenitor for the GRB. Nevertheless, since WR stars are
the most popular candidate progenitors for long GRBs
\citep{1993ApJ...405..273W}, we consult the Galactic Wolf Rayet
Catalogue\footnote{\url{http://pacrowther.staff.shef.ac.uk/WRcat/}}
(v1.25, Aug 2020; \citealt{2015MNRAS.447.2322R}) to estimate the binarity
fraction of WR stars in the Milky Way (MW). The catalog contains 667
Galactic WR stars, among which there are 103 WR stars in binary, yielding a
fraction around 15\%. Due to the usage of a different and updated database,
our result is lower than those in the previous work \citep[e.g., 39\%;][]
{2001NewAR..45..135V}, still being a significant number and indicating that
a large amount of WR stars are in binary systems. Moreover, the binary
fraction of WR stars in the Large Magellanic Cloud (LMC) is found to be 16\%
\citep[or 36\% with the unconfirmed candidates being taken into account;][]
{2014A&A...565A..27H}, and 40\% in the Small Magellanic Cloud \citep[SMC;][]
{2016A&A...591A..22S}. The above facts suggest that in the local universe,
many WR stars can be found in binary systems for MW-like, LMC-like, and
SMC-like metallicity. Assuming all WR stars collapse and lead to GRB events,
we can reasonably conclude that the fraction that a GRB happens in a binary
system, $f_\mathrm{bin}$, is not a small number and can be as high as 50\%
\citep{2009AJ....137.3358M,2015MNRAS.453.1458L}. Considering that not all
GRBs are collapsar GRBs, for simplicity, here we use an estimated
$f_\mathrm{bin}\sim0.2$ to carry out our calculations in this paper.

\begin{figure*}[!t]
 \includegraphics[width=\textwidth]{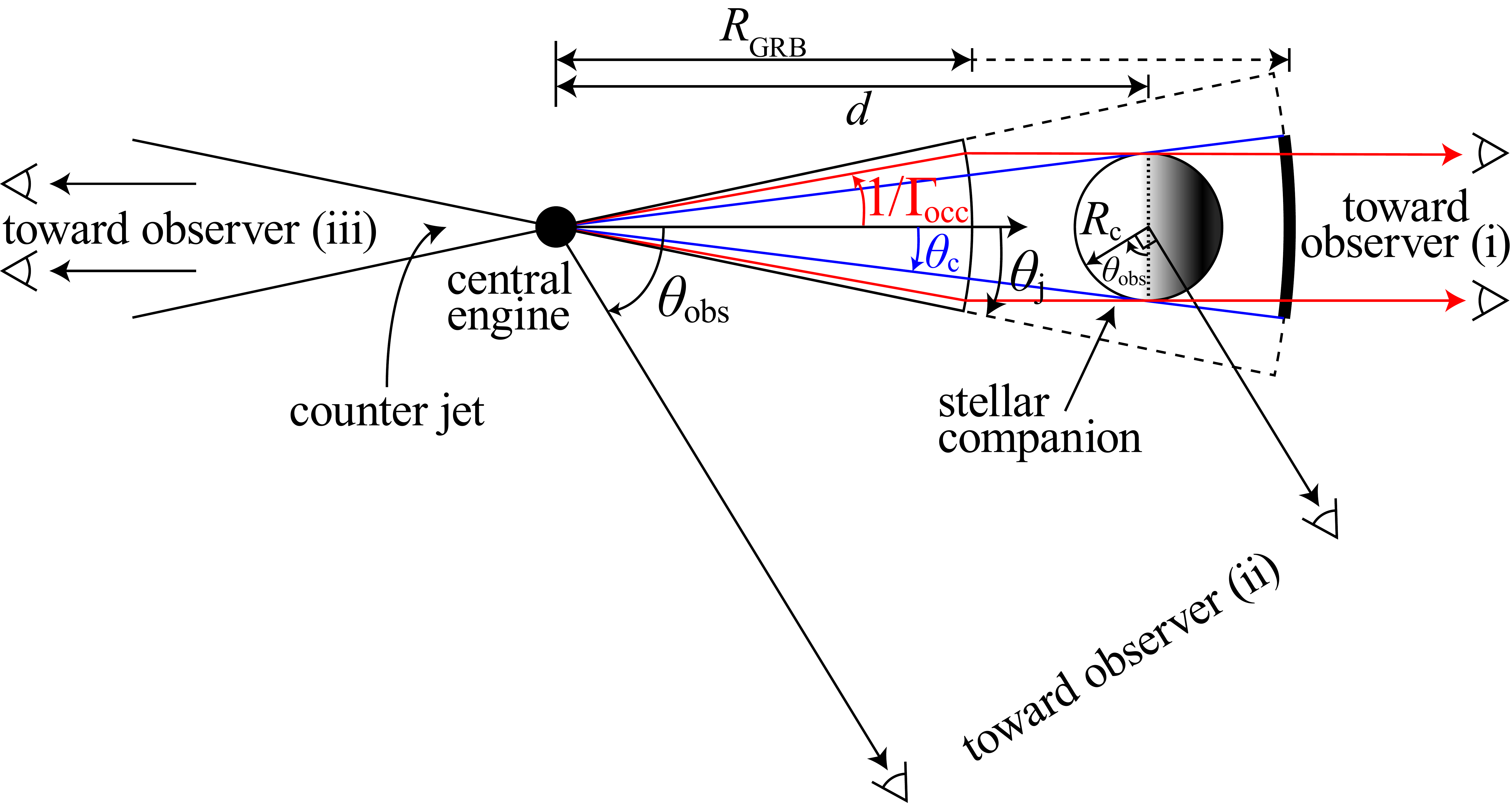}
 \caption{A not-to-scale sketch of the binary geometry with the stellar
 companion located inside the jet opening angle. Sectors with thin solid and
 dashed boundaries in black represent jets with emission sites within and
 beyond the companion, respectively. The thick solid black curve indicates
 the section occulted by the companion. Red and blue lines correspondingly
 indicate the definitions of $\Gamma_\mathrm{occ}$ and $\theta_\mathrm{c}$.
 For convenience, the companion is placed at the exact center of the jet.
 The dotted line divides the companion into a left irradiated hemisphere and
 a right nonirradiated hemisphere. For the observer, three cases are
 illustrated: (i) an on-axis observer, (ii) an observer with a viewing angle
 of $\theta_\mathrm{obs}$, and (iii) an observer located inside the counter
 jet. \label{fig:geometry}}
\end{figure*}

A crucial parameter in the binary system is the separation ($d$) between the
stellar companion and the GRB central engine. Some early studies
\citep[e.g.,][]{1967AcA....17..355P} suggested that WR stars can form
through close binary evolution, while recent observations show that most
Galactic WR binaries seem to have orbital periods longer than 100 days
\citep{2020A&A...641A..26D}, corresponding to $d>1.7\times10^{13}\,$cm for
20 M$_\odot$ circular binaries. An extreme example is the Apep system, which
has an orbital period possibly longer than $10^4$ years \citep
{2019NatAs...3...82C}. Indeed, \citet{2005A&A...435..247P} claimed that the
interacting binary progenitor model is necessary to spin up the WR star core
to produce a long GRB. Nevertheless, whether the binary is interacting or
not is no longer critical when the magnetic torque is involved \citep
{2005A&A...435..247P,2005ApJ...626..350H}. Due to these uncertainties, we
take $d$ as a free parameter. We mainly study the case that $d$ is larger
than the GRB emission radius ($R_\mathrm{GRB}$), but will also briefly
discuss the case of $d<R_\mathrm{GRB}$ (\S \ref{sec:case4}--\ref{sec:case6}).

As illustrated in Figure~\ref{fig:geometry}, we assume that the stellar
companion locates within the GRB jet opening angle, $\theta_\mathrm{j}$;
otherwise, the observational effect will be insignificant. This is
theoretically possible \citep{2018ApJS..237...26H,2020A&A...642A.212J}.
In addition, we assume a random direction of the stellar companion
when determining the observed event rates of such binary systems.

On the other hand, the observing angle (i.e., the angle between our line of
sight and the jet direction) is a key parameter in our framework. As
discussed below, an on-axis observer will see a dimmer GRB due to the
occultation by the stellar companion (when $d<R_\mathrm{GRB}$), whereas an
off-axis observer will see a reflected and delayed GRB even though the
original GRB is missed due to a large observing angle.

The framework we set up in this study involves a highly beamed central
engine and a stellar companion. Besides a WR star, the progenitor of such a
GRB central engine can, in principle, also be a compact star merger (e.g.,
two neutron stars or a neutron star and a black hole). In this case, we have
a hierarchical trinary system before the merge phase, which is quite
possible in dense star clusters \citep{2020ApJ...900...16F}. Of course, some
additional effects, such as a large kick velocity \citep[up to $\sim1000\,
\mathrm{km\,s^{-1}}$ for binary black hole mergers; e.g,][]
{2004ApJ...607L...5F,2004ApJ...607L...9M,2007PhRvL..98i1101G,
2010ApJ...719.1427V} caused by gravitational wave radiation
\citep{1983MNRAS.203.1049F} may be unignorable, yet they are beyond the
scope of this paper. In any case, the framework and theoretical calculations
in this study can also apply to merger-origin short GRBs as long as the
central engine and binary configuration are similar.

\section{Occultation and Reflection Effects} \label{sec:obs}

As discussed above, we only consider the situations when the stellar
companion locates within the opening angle of the GRB jet. With the
uncertainties of the distance between the GRB central engine and the
companion as well as the observer's angle, we need to consider six
different cases of the system configuration (Table~\ref{tab:case}),
which will be studied respectively in the following.

\begin{deluxetable}{cClc}
\tablecaption{Six Different Cases of Binary Geometry \label{tab:case}}
\tablewidth{0pt}
\tablehead{
\colhead{Case} & \colhead{$d\text{ vs. }R_\mathrm{GRB}$} & \colhead{Observer} & \colhead{Sections}}
\startdata
 I    & d\geqslant R_\mathrm{GRB}   & on-axis         & \S\ref{sec:case1} \\
 II   & d\geqslant R_\mathrm{GRB}   & off-axis        & \S\ref{sec:case2} \\
 III  & d\geqslant R_\mathrm{GRB}   & counter-on-axis & \S\ref{sec:case3} \\
 IV   & d<R_\mathrm{GRB}            & on-axis         & \S\ref{sec:case4} \\
 V    & d<R_\mathrm{GRB}            & off-axis        & \S\ref{sec:case5} \\
 VI   & d<R_\mathrm{GRB}            & counter-on-axis & \S\ref{sec:case6} \\
\enddata
\end{deluxetable}

\subsection{Case \textit{I} } \label{sec:case1}

In this case, the GRB emission radius ($R_\mathrm{GRB}$) is less than the
separation between the GRB central engine and the companion, and the
observer is on-axis (observer i in Figure~\ref{fig:geometry}). Depending on
the blockage portion of the jet, the observer will see either a dimmer GRB
or completely miss it.

For simplicity, we assume a uniform jet with a bulk Lorentz factor of
$\Gamma$. Due to the relativistic beaming effect, only those photons emitted
within a cone of $1/\Gamma$ can be observed. Thus the occultation
calculation should be based on the comparison between $1/\Gamma$ and
$R_\mathrm{c}/R_\mathrm{GRB}$, the latter defining the maximal possible
blockage by the companion.

In order to calculate the condition for a total occultation, it is
convenient to define a critical Lorentz factor of
$\Gamma_\mathrm{occ}\equiv R_\mathrm{GRB}/R_\mathrm{c}$. Considering the GRB
framework, $\Gamma_\mathrm{occ}$ can be a few hundred. Taking
$\Gamma_\mathrm{occ}=250$ as an example, we can further calculate its
corresponding scales of the companion and the GRB radius to be
$R_\mathrm{c}=67\,\mathrm{R_\odot}$ and $R_\mathrm{GRB}=10^{15}\,$cm, with
the latter being a typical GRB emission radius for a Poynting flux outflow
\citep{2011ApJ...726...90Z} dominated ejecta. In fact, the initial Lorentz
factor of some GRBs can indeed exceed 250 \citep[e.g.,][]
{2010ApJ...725.2209L,2020ApJ...903L..26H,2021ApJ...909L...3Z}. Therefore,
our calculation suggests that the stellar companion can block the whole GRB
prompt emission if $\Gamma$ does not decrease much in the prompt emission
phase. The prompt emission is unobservable, but the afterglow, with much
smaller values of $\Gamma$, can still be observed as an orphan afterglow
\citep[e.g.,][]{1997ApJ...487L...1R,2002MNRAS.332..735H}.

\begin{figure}
  \epsscale{1.175}
  \plotone{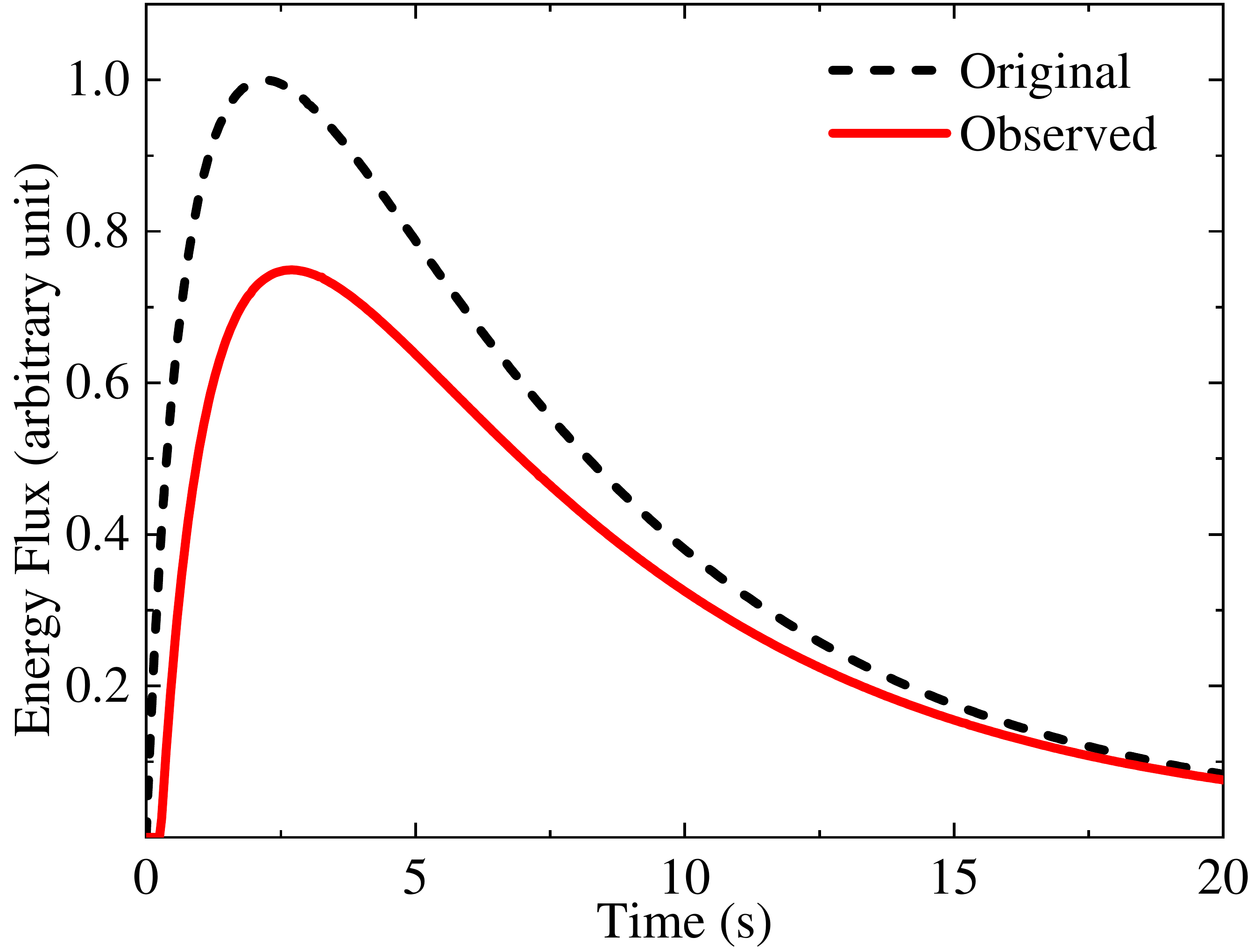}
  \caption{Original (dashed curve) and observed post-occultation (solid
  curve) light curves of a typical GRB. The light curves are normalized so
  that the peak flux of the original light curve equals to unity.
  \label{fig:lc}}
\end{figure}

On the other hand, the Lorentz factor of GRB ejecta can vary in a broad
range \citep[e.g.,][]{2010ApJ...725.2209L} and is not necessarily identical
to $\Gamma_\mathrm{occ}$. So it is more likely that a partial occultation
can occur when $\Gamma<\Gamma_\mathrm{occ}$. In this case, the GRB prompt
emission still can be observed but at a much-reduced level. To study such
occulted GRB light curves, we must include the curvature effect of the
relativistic spherical shell \citep[e.g.,][]{2002ApJ...578..290R,
2003ApJ...596..389K,2004ApJ...614..284D,2015ApJ...808...33U}. For
simplicity, we ignore the details of the radiation mechanisms \citep
{2018ApJS..234....3G} but only consider the pure geometric effect. Following
\citet{2002ApJ...578..290R}, we assume that the relativistic spherical shell
emits an exponential intrinsic pulse in the form of $j(t)\propto\mathrm{exp}
[-t/(5\,\mathrm{s})]$ in its comoving frame. To compare with the total
occultation case, we take $R_\mathrm{c}=67\,\mathrm{R_\odot}$ and $R_\mathrm
{GRB}=10^{15}\,$cm, but a lower Lorentz factor $\Gamma=125$. For a simple
demonstration, we further assume that the stellar companion locates at the
center within the jet opening angle thus it blocks the central $1/\Gamma$
cone of the jet. The partially occultation light curve is then calculated as
\begin{equation}
 I(t)=\int^{+\infty}_{\tau_\mathrm{ang}/\Gamma_\mathrm{occ}^2}\frac{j(t-x)\,
 \mathrm{d}x}{(1+x\Gamma^2/\tau_\mathrm{ang})^2}, \label{eq:occ}
\end{equation}
where $\tau_\mathrm{ang}\equiv R_\mathrm{GRB}/(2c)$ and $c$ is the speed
of light. The original light curve in the observer frame can also be calculated
through Equation~\ref{eq:occ} by setting $\Gamma_\mathrm{occ}\to\infty$.

The results are presented in Figure~\ref{fig:lc}. One can see that the
observed partially occulted light curve generally mimics the original one,
but with a $\sim 25$\% decrease in the flux. At late times, the observed
flux is dominated by the high-latitude emission. Thus, the observed
late-time light curve almost coincides with the original one.

The event rate of an occultation (both partial and total) can be estimated as
\begin{equation}
 \mathcal{R}_\mathrm{occ}\sim\mathcal{R}_\mathrm{obs}f_\mathrm{bin}f_{d
 \geqslant R_\mathrm{GRB}}\left(1/\Gamma+1/\Gamma_\mathrm{occ}\right)^2/4,
\end{equation}
where $\mathcal{R}_\mathrm{obs}$ is the event rate of GRBs and $f_{d
\geqslant R_\mathrm{GRB}}$ is the possibility that $d\geqslant R_\mathrm
{GRB}$. Recalling that most Galactic WR binaries have a relatively long
orbital period, we conservatively estimate $f_{d\geqslant R_\mathrm{GRB}}$
as 0.2. Taking $\Gamma=125$ and $\Gamma_\mathrm{occ}=250$, we finally
estimate the possibility of a GRB being occulted as
$\mathcal{R}_\mathrm{occ}/\mathcal{R}_\mathrm{obs}\sim1.4\times10^{-6}$.

\subsection{Case \textit{II} } \label{sec:case2}

In this case, the GRB emission radius is less than the separation between
the GRB central engine and the companion. The observer is off-axis with a
viewing angle as $\theta_\mathrm{obs}$, as illustrated in Figure~\ref
{fig:geometry} (observer ii). The observer will miss the original GRB
emission but see the reflected radiation by the companion. Note that the
observer may also be able to see the original GRB as well as its reflection
when the line of sight is near the edge of the GRB jet, especially for a
structured jet.

For a GRB with an isotropic energy $E_\mathrm{iso}$, the stellar companion
receives a total energy input of
\begin{equation}
 E_\mathrm{i,tot}\simeq E_\mathrm{iso}\theta^2_\mathrm{c}/4,
\end{equation}
where $\theta_\mathrm{c}\equiv R_\mathrm{c}/d$ is the angular radius of the
companion measured from the GRB central engine. The observer can only see a
fraction ($\theta_\mathrm{obs}/\uppi$) of the irradiated hemisphere.

Considering the projection effect, the effective energy input integrated
over the whole observable reflected energy from irradiated spherical lune
can be derived as
\begin{equation}
 E_\mathrm{i}\simeq\alpha E_\mathrm{iso}\theta^2_\mathrm{c}
 \left(1-\cos\theta_\mathrm{obs}\right)/8.\label{eq:Einj}
\end{equation}
where $\alpha$ is a Bond albedo which defines the reflection efficiency.

Since the incident photons are in gamma-ray energy and the companion surface
is relatively cold, the reflection is mainly through the Compton scattering
process of the electrons \citep{1988ApJ...331..939W,1991ApJ...370..341D}.
The spectrum of the scatted gamma-rays spectrum can be calculated by
employing the Green's function $G(\nu',\nu)$ derived by \citet
{1988ApJ...331..939W}, which is applicable in our case since $h\nu\ll66\,
\mathrm{MeV}$ hence pair production is negligible. The reflected spectrum is
\begin{equation}
 F_{\nu}^\mathrm{r}(\nu)=\int G(\nu',\nu)F_{\nu}^\mathrm{i}(\nu')\,
 \mathrm{d}\nu',
\end{equation}
where $F_{\nu}^{\mathrm{i}}(\nu')$ denotes the incident spectrum, which, in
our case, is a Band function \citep[BF;][]{1993ApJ...413..281B} or a cutoff
power law (CPL; \citealp{2006Natur.442.1008C}), with
$E_\mathrm{peak}$, the lower and higher energy indices
set to 1\,MeV, $-1$, and $-2.5$ (for BF only) respectively.

\begin{figure}
 \epsscale{1.175}
 \plotone{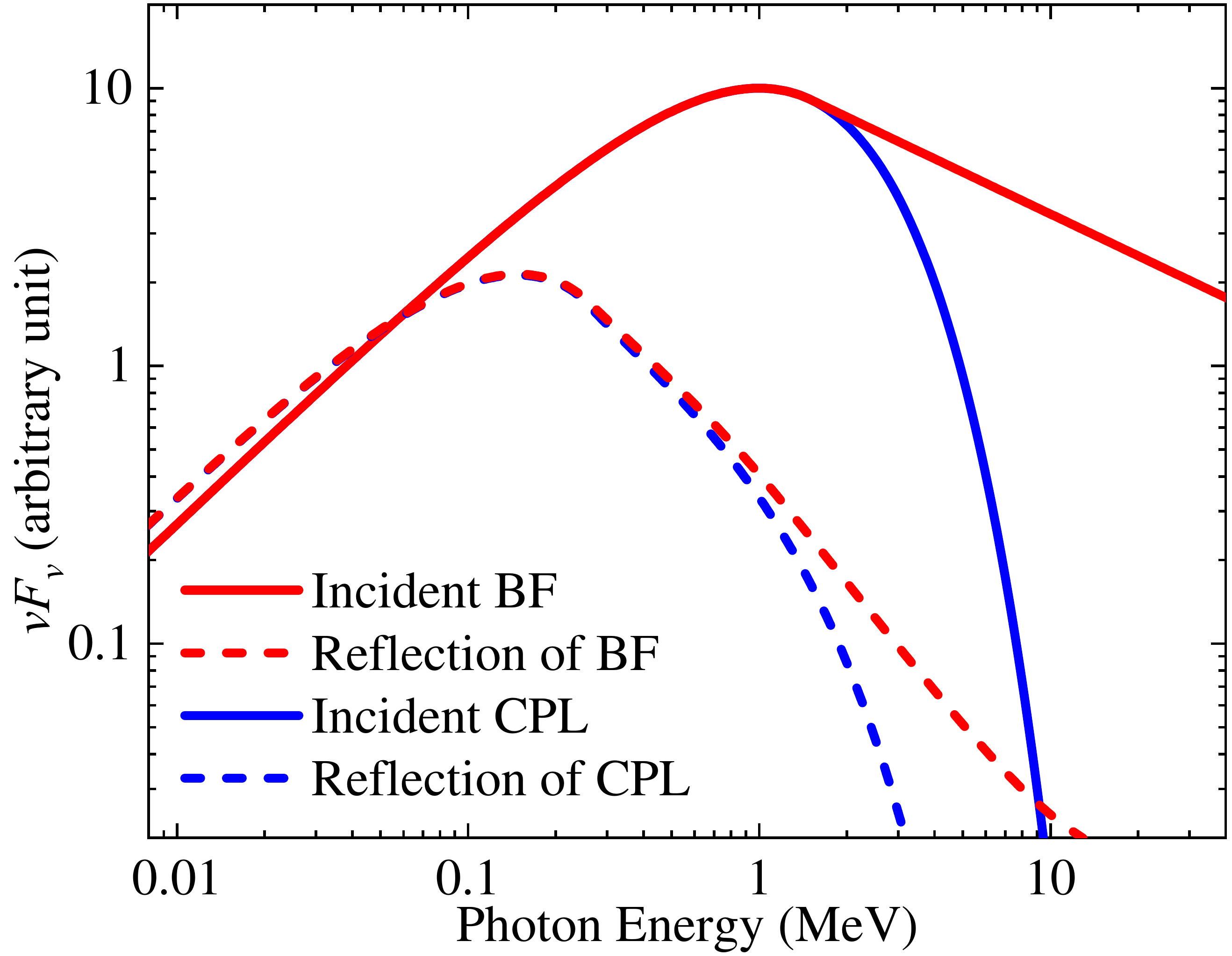}
 \caption{Energy spectra of the incident (solid curves) and reflected
 (dashed curves) emission. The incident spectrum is assumed to be a Band
 function (BF, red curves) or a cutoff power-law function (CPL, blue
 curves). The spectra are normalized to have a peak flux of $\nu F_\nu=10$
 for the incident photons, where $F_\nu$ is the specific flux density at
 frequency $\nu$. \label{fig:spec}}
\end{figure}

The results are shown in Figure~\ref{fig:spec}. Our calculation shows that
the energy flux of the reflected spectrum decreases significantly at higher
energies ($>E_\mathrm{peak}$) and increases slightly at lower energies.
Moreover, the peak photon energy drops by almost one order of magnitude to
143\,keV. We also note that the spectrum shape of the reflected emission
becomes significantly steeper at the high-energy regime.

The reflection efficiency, $\alpha$, can be estimated as the fraction of the
total fluence of reflection and incident flux integrated over the observed
energy range. For a typical BF as show in Figure~\ref{fig:spec}, $\alpha$ is
$\sim 0.33$ for 8\,keV--1\,MeV for \textit{Fermi}-GBM band
\citep{2009ApJ...702..791M}, and is $\sim 0.90$ for 15--150\,keV
\textit{Swift}-BAT band \citep{2005SSRv..120..143B}.

Assuming that the reflected emission is isotropic, the fluence observed by a
distant observer can be calculated by Equation~\ref{eq:Einj} as
\begin{align}
 S_\mathrm{8\,keV-1\,MeV}\sim10^{-8}&\frac{\alpha}{0.2}\frac{E_\mathrm{iso}}
 {10^{54}\,\mathrm{erg}}\left(\frac{R_\mathrm{c}}{67\,\mathrm{R_\odot}}\frac
 {10^{15}\,\mathrm{cm}}{d}\frac{1\,\mathrm{Gpc}}{D_\mathrm{L}}\right)^2\notag
 \\&{}\frac{1+z}{k}\left(1-\cos\theta_\mathrm{obs}\right)
 \,\mathrm{erg\,cm^{-2}},\label{eq:fluence}
\end{align}
where $D_\mathrm{L}$ is the luminosity distance of the GRB (with
$D_\mathrm{L}=1\,$Gpc corresponding to a redshift of $z=0.2$), and $k$ is
the $k$-correction factor. Giving the fact that sensitivity threshold of
\textit{Fermi}-GBM detector is about $\sim10^{-8}\mathrm{erg\,cm^{-2}}$
\citep{2018ApJ...862..152K} in the same energy range, we conclude such
reflected GRB emission is observable under appropriated parameters,
especially for observers at low redshift (e.g., $z\lesssim0.2$). Since the
spectrum (Figure~\ref{fig:spec}) and light curve (see discussion below)
shapes of the reflected emission is very similar to the original incident
GRB, a distant observer may define the reflected emission as a faint GRB
with isotropic energy of
\begin{equation}
 E_\mathrm{iso}^\mathrm{r}\sim10^{48}\frac{\alpha}{0.2}\frac{E_\mathrm{iso}}
 {10^{54}\,\mathrm{erg}}\left(\frac{R_\mathrm{c}}{67\,\mathrm{R_\odot}}\frac
 {10^{15}\,\mathrm{cm}}{d}\right)^2\left(1-\cos\theta_\mathrm{obs}\right)
 \,\mathrm{erg}.\label{eq:Eisoref}
\end{equation}

\begin{figure}
 \epsscale{1.175}
 \plotone{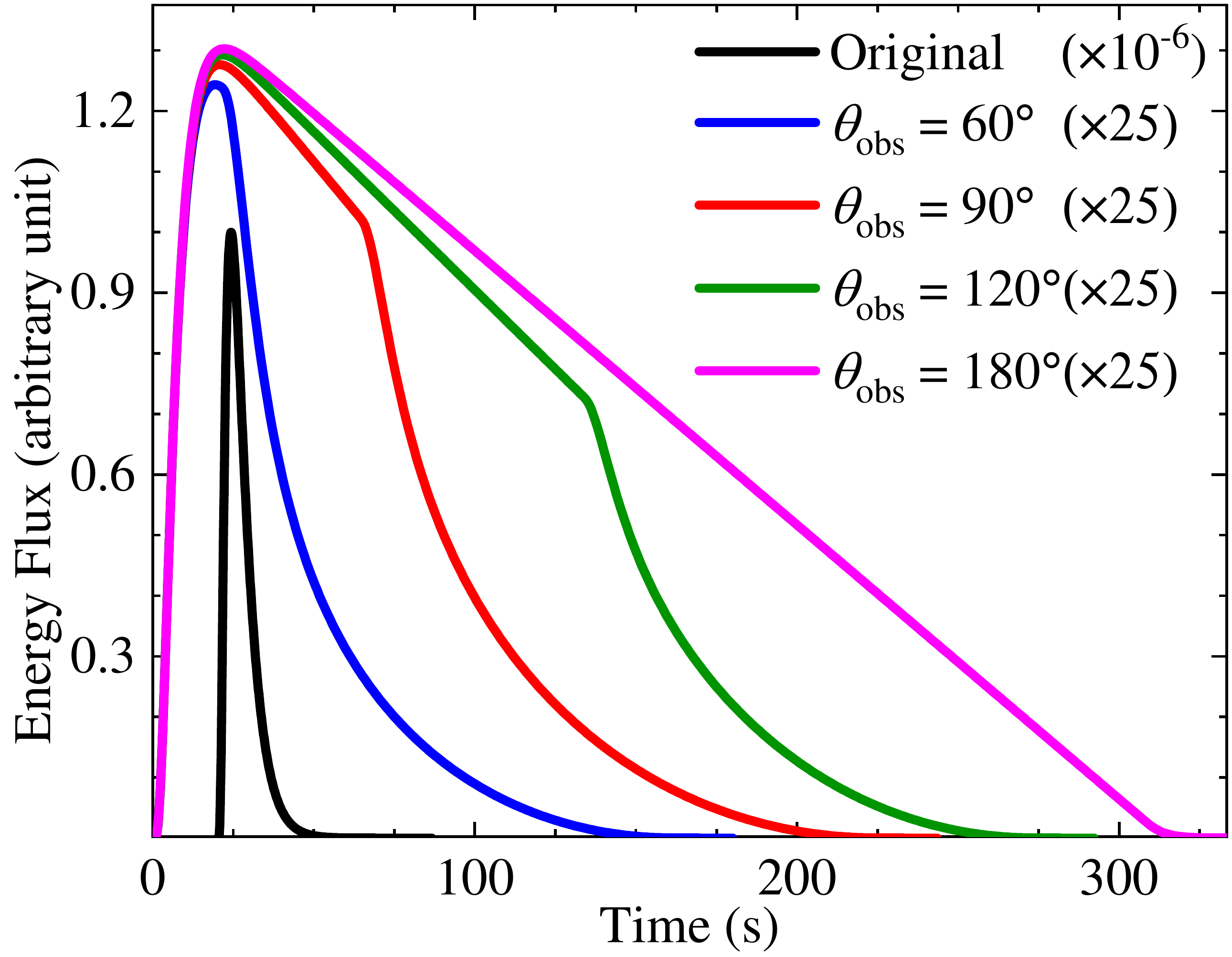}
 \caption{Light curves of the reflected emission produced by a
 $R_\mathrm{c}=67\,\mathrm{R_\odot}$ companion viewed at four different
 angles, i.e., $60^\circ$, $90^\circ$, $120^\circ$, and $180^\circ$ (blue,
 red, green, and magenta curves, respectively). The \textit{y}-axis is
 normalized so that the original GRB has a peak flux of $10^6$. In this
 figure, for better illustration, the original GRB (black curve) has been
 plot after multiplying a factor of $10^{-6}$, and it is also right-shifted
 for 20\,s. \label{fig:broaden}}
\end{figure}

The light curve of the reflected emission can be further calculated by
considering the high latitude effect of the stellar companion. It takes
different times for photons to hit different places of the companion surface
and to be reflected toward the observer. As a result, the ``reflected'' GRB
will be broadened to a time scale of $R_\mathrm{c}/c\sim 2.3\,\mathrm{s}
\times R_\mathrm{c}/\mathrm{R_\odot}$. By considering the projection effect,
the exact reflection light curve depends on the viewing angle and can be
calculated by
\begin{equation}
 I^\mathrm{r}(t)=\frac{2}{\uppi}\int_0^{\frac{\uppi}{2}}\int_{\frac{\uppi}
 {2}-\theta_\mathrm{obs}}^{\frac{\uppi}{2}}I^\mathrm{i}\left[t-t'\left(R_
 \mathrm{c},\theta_\mathrm{obs},\theta,\delta\right)\right]\sin^2\theta
 \cos\delta\,\mathrm{d}\theta\,\mathrm{d}\delta,
\end{equation}
where $t'\left(R_\mathrm{c},\theta_\mathrm{obs},\theta,\delta\right)$ is the
extra travel time of photons hitting $(\theta,\delta)$ point on the
companion surface, and $I^\mathrm{i}(t)$ is the incident light curve. In
Figure~\ref{fig:broaden}, we plot the light curves of an original GRB and
its reflection emissions viewed at different angles. The companion radius is
taken as $67\,\mathrm{R_\odot}$. We can see that the reflection light curves
are broader for larger viewing angles, yet the peak flux is essentially a
constant. Moreover, the light curves first decline linearly and break to a
convex shape for intermediate viewing angles (e.g., $90^\circ$ and
$120^\circ$). This break is due to the equal-time surface effect. When the
equal-time surface reaches the edge of the observable irradiated spherical
lune, a break occurs. For a small $\theta_\mathrm{obs}$ (e.g., $60^\circ$),
the linear decline phase is too short ($\sim 21\,\mathrm{s}$) to be observed
in the light curve. On the other hand, for a large viewing angle such as
$\theta_\mathrm{obs}=180^\circ$, the break happens at very late stages, and
it is not visible on the plot.

\begin{figure}
 \epsscale{1.175}
 \plotone{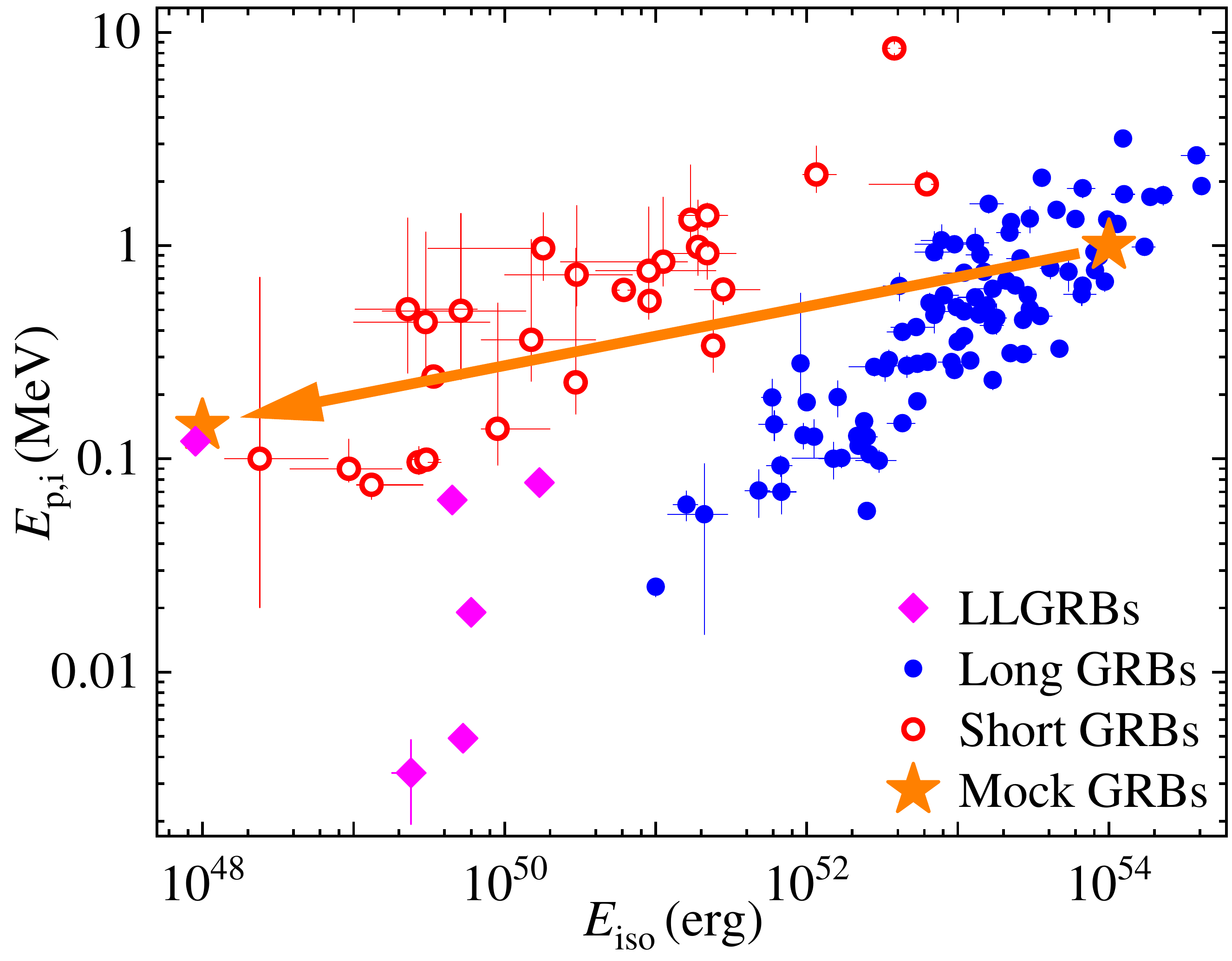}
 \caption{$E_\mathrm{p,i}$ vs. $E_\mathrm{iso}$ for low-luminosity GRBs
 (LLGRBs, diamonds; \citealp{2018pgrb.book.....Z}), long GRBs (filled
 circles; \citealp{2018NatAs...2...69Z}), and short GRBs (open circles;
 \citealp{2018NatAs...2...69Z}). A mock GRB and its reflection GRB are shown
 in star symbols, with an arrow pointing from the original mock GRB to the
 reflected one. All error bars represent 1-$\sigma$ uncertainties.
 \label{fig:amati}}
\end{figure}

Our results shed some light on the origin of low-luminosity GRBs (LLGRBs).
LLGRBs are usually considered to form a distinct population of long GRBs
\citep[e.g.,][]{2007ApJ...662.1111L,2009MNRAS.392...91V}. They have a very
low luminosity and relatively soft spectrum \citep{2015ApJ...807..172N}. The
physical origin of LLGRBs is still unresolved. Unlike previous attempts,
which involve a special change of the GRB jet radiation or dynamics (e.g.,
models involving a chocked jet, \citealp{2012ApJ...747...88N,
2012ApJ...749..110B,2015ApJ...807..172N}; a long-lived central engine
powering a mildly relativistic jet, \citealp{2016MNRAS.460.1680I}; or the
collision between the relativistic ejecta and circumstellar medium, \citealp
{2019ApJ...870...38S}), our model, by only introducing the stellar
companion's reflection effect, can naturally reproduce the observed LLGRBs.
This is manifested in Figure~\ref{fig:amati}, where we over-plotted a mock
GRB with $E_\mathrm{iso}=10^{54}\,$erg and rest-frame peak energy $E_\mathrm
{p,i}=1\,$MeV and its reflected burst together with a large sample of short
and long GRBs, as well as the previously identified LLGRBs
\citep{2018pgrb.book.....Z}, on the Amati relation plot
\citep{2002A&A...390...81A}. Interestingly, the reflected GRB shows up as a
weak GRB with $E_\mathrm{p,i}\sim143\,$keV and $E_\mathrm{iso}\sim
10^{48}\,$erg and appears in the LLGRB region, closed to GRB~980425
\citep{2007ApJ...654..385K,2007ApJ...662.1111L,2009MNRAS.392...91V}. In
addition, we also note that the ``reflected'' GRBs usually have a long
duration (see Figure~\ref{fig:broaden}), which is also well consistent with
the observed LLGRBs.

Our model predicts a moderate event rate for low-luminosity GRBs. The
observed event rate can be estimated as
\begin{equation}
 \mathcal{R}_\mathrm{r}=\mathcal{R}_\mathrm{obs}f_\mathrm{bin}
 f_{d\geqslant R_\mathrm{GRB}}.
\end{equation}
Typically, we estimate that the observed ratio of low-luminosity GRBs over
normal long GRBs is $\mathcal{R}_\mathrm{r}/\mathcal{R}_\mathrm{obs}
\sim0.04$, which can explain at least a fraction of the observed LLGRBs
\citep[e.g.,][]{2007ApJ...662.1111L,2009MNRAS.392...91V}.

\subsection{Case \textit{III} } \label{sec:case3}

In this case, the GRB emission radius is less than the separation between
the GRB central engine and the companion, and the observer is on-axis.
However, the companion star is located within the opening angle of the
counter jet, as demonstrated as observer iii in Figure~\ref{fig:geometry}.
It is also equivalent to assign a viewing angle of $\theta_\mathrm{obs}=
\uppi$. The observer will see both the original and the reflected GRBs.

As pointed out by \citet{1991ApJ...370..341D}, the distance between the
companion and the GRB central engine will cause a time delay between the
original GRB and its reflection. Since the GRB jet moves at a highly
relativistic speed, the time delay can be calculated as
\begin{equation}
 \tau=\frac{d}{c}\left(1-\cos\theta_\mathrm{obs}\right)\sim6.7
 \times10^4\frac{d}{10^{15}\,\mathrm{cm}}\,\mathrm{s}. \label{eq:delay}
\end{equation}
The reflected emission may manifest as a soft gamma-ray or hard X-ray bump
superposed on the GRB high-energy afterglow. Combining with multi-messenger
signals such as gravitational waves, the original and reflected GRBs, one
may use Equation~\ref{eq:delay} to further constrain the radiation mechanism
through constraining of the prompt emission site by the time delay of the
different signals \citep[e.g.,][]{2017ApJ...850L..24G,2018PTEP.2018d3E02I,
2018NatCo...9..447Z} if the GRB occurs with a stellar companion.

The observed event rate of such a reflection bump can be estimated as
\begin{equation}
 \mathcal{R}_\mathrm{bump}=\mathcal{R}_\mathrm{obs}f_\mathrm{bin}
 f_{d\geqslant R_\mathrm{GRB}}\theta_\mathrm{j}^2/4.
\end{equation}
Taking the jet opening angle as a typical value of 10 degrees, we get
$\mathcal{R}_\mathrm{bump}/\mathcal{R}_\mathrm{obs}\sim3\times10^{-4}$.

\subsection{Case \textit{IV} } \label{sec:case4}

In this case, the GRB emission radius is larger than the separation between
the GRB central engine and the companion, and the observer is on-axis.
Similar to Case (I), the observer will see a dimmer GRB if the companion
star does not fully occult the GRB but blocks a $\theta_\mathrm{c}=R_\mathrm
{c}/d$ cone of the jet. The observed light curve is similar to Case (I), but
with $\Gamma_\mathrm{occ}$ defined as $\Gamma_\mathrm{occ}=d/R_\mathrm{c}$.

\subsection{Case \textit{V} } \label{sec:case5}

In this case, the GRB emission radius is larger than the separation between
the GRB central engine and the companion, and the observer is off-axis with
a viewing angle of $\theta_\mathrm{obs}$. The observer will miss the
original GRB but will see collision emission between the jet and the
companion, probably causing a shock \citep{2005astro.ph.10192M}. The physics
behind this scenario is that the relativistic jet materials will collide
with the companion, producing a very strong blastwave. The collision will
convert the jet kinetic energy into internal energy and generate X-ray
emission \citep{2005astro.ph.10192M}. Such emission will not manifest as a
GRB but may show up like an X-ray transient.

\subsection{Case \textit{VI} } \label{sec:case6}

In this case, the GRB emission radius is larger than the separation between
the GRB central engine and the companion, and the observer is on-axis.
However, the companion star is located within the opening angle of the
counter jet. The geometry configuration is similar to Case (III), and the
radiation mechanism is identical to Case (V). Consequently, the observer
will see the original GRB, followed by a delayed X-ray transient produced by
the relativistic counter jet interacting with the companion star. The X-ray
transient may show up as an X-ray flare
\citep[e.g.,][]{2005Sci...309.1833B,2006ApJ...641L..89W,2010MNRAS.406.2113C,
2017ApJ...841L..15G,2018ApJ...862..115G}, appearing at a time of several
hundreds of seconds after the burst \citep{2016ApJS..224...20Y} on top of
the X-ray afterglows. Although this collision scenario can hardly
produce multiple X-ray flares in a GRB, it may apply to at least a fraction
of GRBs with only one single X-ray flare detected.

\begin{figure*}[!t]
  \includegraphics[width=\textwidth]{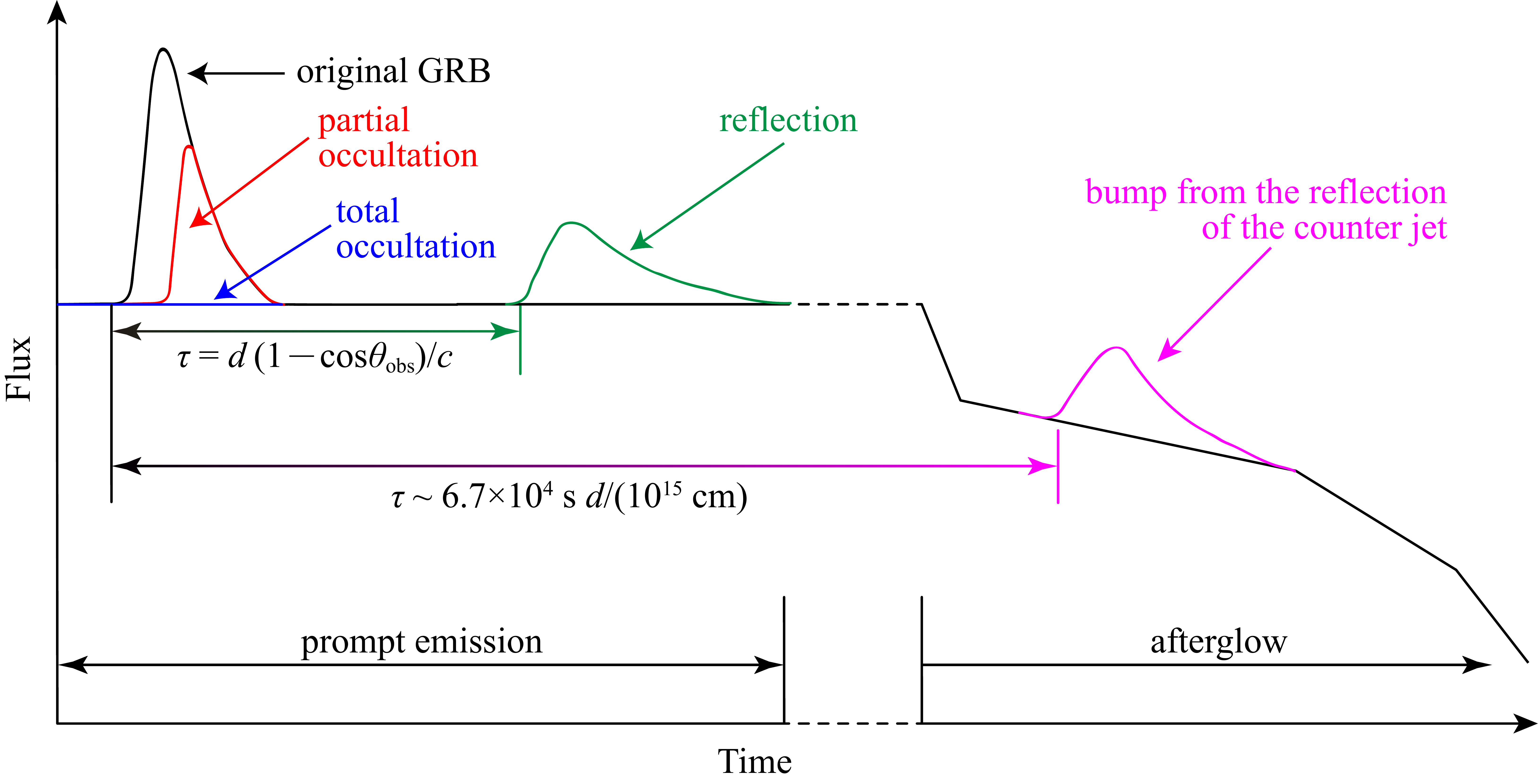}
  \caption{An overall illustration of the various phenomena possibly
  associated with a GRB that happens in a binary system. The original light
  curves (solid black curves) of the prompt emission and X-ray afterglow of
  the GRB are schematically connected by a dashed line. Light curves of the
  total occultation, partial occultation, reflection, and bumps from the
  reflection of the counter jet are shown by blue, red, green, and magenta
  solid curves. \label{fig:overall}}
\end{figure*}

Besides generating X-ray emission, the collision can also heat the
companion star's atmosphere and expand its outer layer. Consequently, an
optical/infrared transient may occur after the GRB \citep{2005astro.ph.10192M}.
The typical energy of such transient can reach as high as $\sim10^{48}\,$erg
\citep{2005astro.ph.10192M} and may be registered as an early optical flare
\citep{2013ApJ...774....2S}. On the other hand, such a transient is unlikely
to be mistaken for a supernova bump because a supernova bump usually shows up
about two weeks after the GRB
\citep[e.g.,][]{1998Natur.395..670G,2003Natur.423..847H}, which is much
longer than the typical time delay in our scenario (i.e., around one day
by Equation~\ref{eq:delay}).

\section{Summary and Discussion} \label{sec:conclude}

In this paper, we performed a comprehensive study on the effect caused by
the stellar companion when a GRB happens in a binary system. Our results
are sketched in Figure~\ref{fig:overall}, which illustrates the two main
effects, namely occultation and delayed reflection. In addition, we also
pointed out that the collision between the jet materials and the stellar
companion will generate X-ray emission if the stellar companion is
located inside the GRB beam, which is observable by off-axis or
counter-on-axis observers.

We note that some additional configurations of the binary system were not
included in our analysis. For example, when the companion is partially or
entirely outside the jet opening angle \citep{2015MNRAS.453.1458L}, the
observer can see its inverse-Compton scattered emission, with(without) the
original GRB if the observer is on(off)-axis.

GRBs are generally believed to be highly beamed \citep{1997ApJ...487L...1R,
1999MNRAS.309..513H,2000A&A...355L..43H,2000MNRAS.316..943H,
2000ApJ...543...90H,2001ApJ...562L..55F,2015ApJ...815..102F,
2018ApJ...859..160W}. Therefore, reflection is expected to be more likely
than occultation since an occultation requires the observer to be on-axis.
The observational features discussed in this paper can be tested by
observations and used to probe the properties of the GRB systems.

\begin{acknowledgments}
 We thank the anonymous referee for
 valuable suggestions that lead to an overall improvement of this study.
 Z.-C. Z. gratefully acknowledges Yong Shao for helpful discussions on WR
 stars and Xiangyu Wang for assistance with GRB data. BBZ thanks B. Zhang
 for helpful discussions on the overall science of this study. This work is
 supported by National SKA Program of China No. 2020SKA0120300, by the
 National Natural Science Foundation of China (Grant Nos. 11873030,
 12041306, U1938201, U1831135, 11833003, U2038105). B.B.Z acknowledges
 support by Fundamental Research Funds for the Central Universities
 (14380046), the science research grants from the China Manned 
 Space Project with NO.CMS-CSST-2021-B11, and the Program for Innovative Talents,
 Entrepreneur in Jiangsu. We acknowledge the use of public data from the
 Fermi Science Support Center (FSSC).
\end{acknowledgments}

\vspace{1cm}

\software{Astropy \citep{2013A&A...558A..33A,2018AJ....156..123A},
 NumPy \citep{2011CSE....13b..22V,2020Natur.585..357H},
 OriginPro (\url{https://www.originlab.com/}),
 SciPy \citep{2020NatMe..17..261V}
}

\bibliographystyle{aasjournal}

\end{document}